\documentclass[prl,aps,twocolumn,showpacs,amsmath,amssymb,superscriptaddress]{revtex4}

\usepackage{graphicx}
\usepackage{dcolumn}
\usepackage{bm}
\usepackage{natbib}

\begin{document}

\title{Aharonov-Bohm Oscillations Changed by Indirect Interdot Tunneling via Electrodes in Parallel-Coupled Vertical Double Quantum Dots}

\author{T. Hatano}
\thanks{Present Address: JST, ERATO, Hirayama Nuclear Spin Electronics Project, 468-15 Aramaki Aza Aoba, Aoba-ku, Sendai 980-0845, Japan; Electronic address: hatano@ncspin.jst.go.jp}
\affiliation{JST, ICORP, Quantum Spin Information Project, Atsugi-shi, Kanagawa 243-0198, Japan}

\author{T. Kubo}
\affiliation{JST, ICORP, Quantum Spin Information Project, Atsugi-shi, Kanagawa 243-0198, Japan}

\author{Y. Tokura}
\affiliation{JST, ICORP, Quantum Spin Information Project, Atsugi-shi, Kanagawa 243-0198, Japan}
\affiliation{NTT Basic Research Laboratories, Atsugi-shi, Kanagawa 243-0198, Japan}

\author{S. Amaha}
\affiliation{JST, ICORP, Quantum Spin Information Project, Atsugi-shi, Kanagawa 243-0198, Japan}

\author{S. Teraoka}
\affiliation{JST, ICORP, Quantum Spin Information Project, Atsugi-shi, Kanagawa 243-0198, Japan}

\author{S. Tarucha}
\affiliation{JST, ICORP, Quantum Spin Information Project, Atsugi-shi, Kanagawa 243-0198, Japan}
\affiliation{Department of Applied Physics, University of Tokyo, Hongo, Bunkyo-ku, Tokyo 113-8656, Japan}

\date{\today}

\begin{abstract}
Aharonov-Bohm (AB) oscillations are studied for a parallel coupled vertical double quantum dot with a common source and drain electrode. We observe AB oscillations of current via a one-electron bonding state as the ground state and an anti-bonding state as the excited state. 
As the center gate voltage becomes more negative, the oscillation period is clearly halved for both the bonding and antibonding states, and the phase changes by half a period for the antibonding state. 
This result can be explained by a calculation that takes account of the indirect interdot coupling via the two electrodes. 
\end{abstract}

\maketitle

Quantum coherence and the correlation of electrons in semiconductor nanostructures are of primary interest. In the past few years, they have been well characterized for qubit states in single and double quantum dots (DQDs) in the context of quantum information. Before this, the direct observation of quantum coherence in QDs had already been achieved using the Aharonov-Bohm (AB) effect \cite{Aharonov}. 
Conductance or current through an AB ring oscillates periodically with magnetic flux penetrating the ring area. 
The inclusion of a QD in the AB ring causes a phase change in the AB oscillations corresponding to changes in the number of electrons in the dot \cite{Yacoby}. 
The abrupt phase change by the half period was initially observed \cite{Yacoby}, and explained using the reciprocal theorem for a two-terminal measurement \cite{Buttiker}. 

AB oscillations in current flowing through two QDs or DQDs embedded in an AB ring with one QD in each arm were studied theoretically with particular attention to the electron correlation effect \cite{Akera,Izumida}. 
The AB phase of a current flowing through two-electron states in a tunnel coupled DQD is predicted to be different by half a period in the cotunneling regime between the spin singlet state and triplet state due to interdot exchange coupling \cite{Loss}.
In contrast, for one-electron states, it is anticipated that the AB oscillations of the bonding state (BS) are phase changed by half a period from those of the anti-bonding state (AS), reflecting the parity of the electron wavefunctions in the DQD \cite{Kang,Kubo,Tokura}. 

These features have been experimentally studied for parallel coupled lateral DQDs \cite{Holleitner} and vertical DQDs \cite{Hatano4} with a common source and drain electrodes, but without reference to the phase evolution. 
The phase change in the AB oscillation has only recently been observed in association with inelastic cotunneling processes but unrelated to interdot coupling \cite{Sigrist2}. In all previous reports, the effect of interdot coupling was too weak to be resolved in the AB oscillations.

In this paper, we use a laterally coupled vertical DQD with a common source and drain electrodes to study the AB oscillations of current through the one-electron states. 
We first precisely identify the one-electron BS as the ground state and the AS as the excited state in the charging diagram, and then measure the AB oscillations of current through the BS and AS. 
We observe critical changes in the phases and periods of the AB oscillations, depending on the center gate voltages. 
To account for these results, we consider the effect of the extra interdot tunnel coupling through the electrodes ("indirect interdot tunnel coupling").

\begin{figure}
\centering
\includegraphics[width=1\columnwidth]{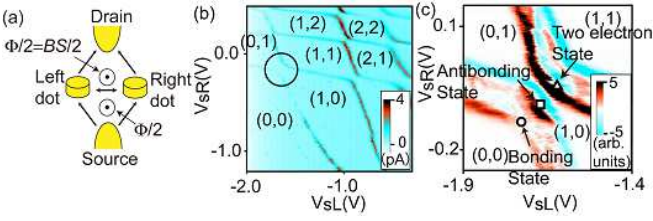}
\caption{
(a) Diagram of current flow ($I_{SD}$) in parallel coupled double quantum dot. The arrows indicate the current flow. 
 (b) Color plot of the current as a function of the left and right side gate voltages, $V_{sL}$ and $V_{sR}$, at a source drain voltage $V_{SD}$=50 $\mu$V and a center gate voltage $V_c=-1.20$ V. 
($N_L, N_R$) denotes the electron numbers in the left and right dots. 
 (c) Color plots of the average values of the transconductance $(dI_{SD}/dV_{sL}+dI_{SD}/dV_{sR})/\sqrt{2}$ as a function of $V_{sL}$ and $V_{sR}$ at $V_c=-1.20$ V and $V_{SD}$=800 $\mu$V, corresponding to the region within the circle in (b). 
} 
\label{figure2}
\end{figure}

Our device consists of laterally coupled vertical DQDs with four split gates made from a double-barrier heterostructure (DBH)  \cite{epaps,Hatano,Hatano2,Hatanoprb}. 
A current flows vertically through the two parallel coupled QDs as shown in Fig.~\ref{figure2}(a). 

It is usually assumed that electron coherence readily collapses in the electrodes. However, this was not the case in our previous experiment on the same type of DQD device as used here because AB oscillations were indeed observed for the AB ring encircled through the source and drain \cite{Hatano4}. 
This is because the electrodes are adjacent to the thin AlGaAs tunnel barriers just outside the QDs. 
Therefore, we assume for the vertical QDs that the two QDs are coupled via the n-AlGaAs contact layers in a sufficiently short path length to maintain the coherence. 
Here, we introduce the indirect interdot tunnel coupling parameter $\alpha_{s}$ and $\alpha_{d}$ $(|\alpha_{s(d)}| \leqq 1)$, which indicates the degree of coherence in the source and drain, respectively \cite{Kubo}. 
When $\alpha_{s(d)}=0$, the coherence is completely lost; the AB oscillation is not observed. In contrast, the coherence in the electrodes is completely conserved at $|\alpha_{s(d)}|=1$; the visibility of the AB oscillation is at its maximum. Note that the visibility of the AB oscillation is 100\% when the energy level in each dot is aligned. In actual systems, however, such a $|\alpha_{s(d)}|=1$ case is very special and most experimental situations correspond to $|\alpha_{s(d)}| <$ 1. 
In particular, in the measured device, we assume $|\alpha_{s(d)}| \lesssim 0.2 $, as described below \cite{epaps}.  

Then two current loops with more or less the same area, $S/2$, are formed due to the presence of the direct interdot tunnel coupling. 
The magnetic flux, which penetrates the large current loop of the area $S$, including the two small same-sized loops, is defined by $\Phi=BS$ as shown in Fig.~\ref{figure2}(a), where $B$ is an externally applied in-plane magnetic field. Thus, the magnetic flux, which penetrates each small loop, is $\Phi/2$. 
The evaluated AB oscillation period is $\Delta B_0=2\Phi_0/S=0.77$T, where $\Phi_0=h/e$ is the magnetic flux quantum, $h$ is the Planck constant and $e$ the electron charge \cite{epaps}. 
All the transport measurements are performed on a device placed in a dilution refrigerator with a base temperature of 20 mK. 

Figure~\ref{figure2}(b) shows current ($I_{SD}$) peaks or Coulomb peaks evolving with left and right gate voltages, $V_{sL}$ and $V_{sR}$, measured at a bias voltage, $V_{SD}$=50 $\mu$V. The center gate voltage, $V_c$, is fixed at -1.20 V. No Coulomb peaks are observed for $V_{sL}\lesssim -1.7$ V and $V_{sR}\lesssim -0.1$ V, indicating that the DQD is empty. Then, the electron number $N_L(N_R)$ in the left (right) QD increases by one whenever the point in the parameter space in Fig.~\ref{figure2} (b) crosses one of the Coulomb peaks on the right (top). 
Therefore, the charge states with $(N_L, N_R)$ are fixed in each hexagon formed by the Coulomb peaks. 

Figure~\ref{figure2}(c) shows the average value of the transconductance $(dI_{SD}/dV_{sL}+dI_{SD}/dV_{sR})/\sqrt{2}$ as a function of $V_{sL}$ and $V_{sR}$ at $V_c=-1.20$ V and $V_{SD}$=800 $\mu$V. 
The two anti-crossing stripes indicate that the Coulomb peaks are widened by applying a finite $V_{SD}$. The lower left, and upper right stripes correspond to the one-, and two-electron states, respectively. 
The red or black lower border line of each stripe indicates the ground state, and the red or black line inside the lower left stripe is the one-electron excited state \cite{Hatanoprb}. 
For the one-electron stripe, we observe that the ground and excited states anti-cross with each other as a typical feature of the resonance between the BS and AS, which  are indicated by the circle and square in Fig.~\ref{figure2}(c), respectively \cite{Hatanoprb}. 
We estimated an interdot tunnel coupling $2t$ of 0.53 meV. The interdot Coulomb coupling energy $V_{inter}$ is approximately 0.4 meV. 
The tunneling rates from the source and drain electrodes to the DQD, $\Gamma_{s}$ and $\Gamma_{d}$ are estimated to be $\sim 0.5$ $\mu$eV and $\sim 1.5$ $\mu$eV, respectively. 

For the two-electron state, the excited state is not well resolved in Fig.~\ref{figure2}(c), probably because the exchange coupling energy that corresponds to the energy separation between the singlet and triplet states is very small \cite{Hatanoprb}. Therefore, the state indicated by the triangle is the two-electron ground state.

\begin{figure}
\centering
\includegraphics[width=1\columnwidth]{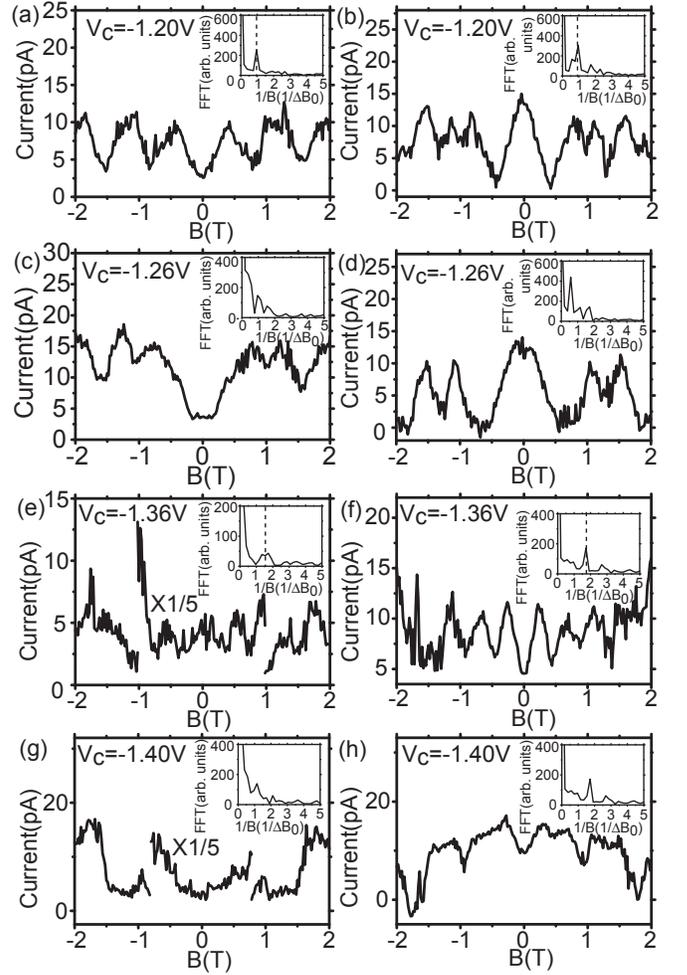}
\caption{
Current flowing through the bonding state (a), (c), (e) and (g) and the anti-bonding state (b), (d), (f) and (h) as a function of $B$ for $V_c=$-1.20, -1.26, -1.36 and -1.40 V and at $V_{SD}=$800, 700, 500 and 500 $\mu$V, respectively. Each inset shows the corresponding fast Fourier transform (FFT) spectrum of the current oscillation, where $\Delta B_0=2\Phi_0/S=0.77$ T.} 
\label{figure3}
\end{figure}

We measured the current at the circle and square points as a function of $B$ for various $V_c$ values. The two points are defined by the charging diagram measurement at $B=$0T (not shown). 
Here, let us consider the BS and AS currents. 
The current at the circle in Fig.~\ref{figure2}(c) flows solely through the BS because only the BS is within the bias window, which is caused by applying $V_{SD}$. 
Consequently, the current is the BS current. 
On the other hand, at the square the current flows through the BS and AS, since both states are in the bias window. Using the BS and AS as the bases, the diagonal elements of a transmission matrix indicate the transmission probabilities of the BS and AS, and the total current is calculated by integrating the transmission probabilities  in the bias window with respect to energy \cite{epaps,Meir}. 
Moreover, the current flowing through the system generally accompanies the mixing component of the BS and AS currents, which corresponds to the higher order tunneling processes comprising both BS and AS. However, these processes, which are proportional to the second order terms of $\alpha_{s(d)}$, can be neglected, since $\alpha_{s(d)}$ is very small in our case \cite{epaps}.
Therefore, we can simply describe the total current as the sum of the BS and AS currents. Then, we can assume that the BS currents at the circle and square are almost the same, because the potential barrier height, i.e. the electron transmission probability, at the circle and square hardly changes. 
Consequently, we subtracted the current measured for the BS from that at the square to extract the AS current.

The currents obtained for the BS, and AS are shown as a function of $B$ in Fig.~\ref{figure3}(a), and (b), respectively, where $V_{SD}$ and $V_{c}$ are the same as those in Fig.~\ref{figure2}(c).  
Both currents oscillate periodically with $B$; these oscillations are AB oscillations.
The fast Fourier transform (FFT) spectra of (a) and (b) in the insets show peaks at 0.81 T, and 0.79 T, respectively. These values are consistent with the calculated value of $\Delta B_0$=0.77 T. 
Therefore, this oscillation corresponds to area $S/2$ rather than area $S$ owing to direct interdot tunnel coupling \cite{Kang,Kubo}.  
There is a good half period contrast between (a) and (b) because there is a current dip at $B=0$ T for BS whereas there is a peak for AS. 
It should be noted that the small dip at around $\pm1$ T in Fig.~\ref{figure3} (b) is probably because the contribution of the BS current is not completely subtracted in the derivation of the AS current. 

The BS and AS currents obtained for $V_c=$-1.26, -1.36 and -1.40 V and at $V_{SD}=$700, 500 and 500 $\mu$V are shown in Fig.~\ref{figure3}(c) to (h), respectively. 
Here $2t$ is estimated to be 0.46, 0.35 and 0.24 meV at $V_c=$-1.26, -1.36, and -1.40 V, respectively. The interdot Coulomb coupling energy $V_{inter}$ is approximately 0.4 meV for the three different $V_c$ values. 
In Fig.~\ref{figure3}(c) and (d) at $V_c=-1.26$ V, the current oscillations are not very periodic but on the whole they are similar to those in Fig.~\ref{figure3}(a) and (b), respectively. 
The FFT spectrum in Fig.~\ref{figure3}(d) shows that the largest peak is slightly different from that in Fig.~\ref{figure3}(b), reflecting the larger peak spacing at $B=0$ T and $\pm 1.1$ T. 

The currents obtained for the more negative $V_c$ values show markedly different features. 
At $V_c=-1.36$ V, the BS current oscillates only weakly with $B$ in Fig.~\ref{figure3}(e). The oscillation period of 0.5 T as identified in the FFT spectrum is approximately half the value obtained for the BS current at $V_c=-1.20$ V (a) and -1.26 V (c). 
In contrast, for the AS current in Fig.~\ref{figure3}(f), we observe clear oscillations with a period of 0.44 T, which is almost the same as that for the BS current in (e). 
The AB oscillation of the AS current in (f) with $V_c=-1.36$ V has a dip at $B=0$ T, whereas it has a peak at $V_c=-1.20$ V (b) and $V_c=-1.26$ V (d).  On the other hand, that of the BS current in (e) for $V_c=-1.36$ V seems to leave a small peak at $B=0$ T, judging from the magnified plot. However, because the BS current is smaller than 1~pA for $|B|<1$ T and comparable to the noise level, it is difficult to resolve the AB oscillation at around $B=0$ T. 

The AB oscillations are significantly damped for both BS and AS, when $V_c$ becomes even more negative at -1.40 V in Fig.~\ref{figure3}(g) and (h). The features of the AB oscillations appear similar to those at $V_c=-1.36$ V, although the AB oscillation of the BS is not very clear. 
The application of a more negative voltage to the center gate voltage weakens the direct interdot tunnel coupling. Moreover, it increases the dephasing in the electrodes, because of the increase in the effective path length in the electrode to connect the two dots. The latter could be the reason for the reduced visibility of the AB oscillations at $V_c=-1.36$ V and -1.40 V in Figs.~\ref{figure3}(e) to (h). 

However, possible changes in the distance between the two dots and the direct interdot tunnel coupling and in the dephasing caused by applying a much more negative $V_c$ cannot account for either the abrupt changes in the phase or the period of AB oscillations with $V_c$ as observed in Fig.~\ref{figure3}. 
Here, we consider that the interference effect induced by the indirect interdot tunnel coupling via the electrodes \cite{epaps} accounts for the observed features. 

First, let us simply analyze the currents flowing through the BS and AS, when the bias voltage is so large that the BS and AS are in the bias window. 
We define the tunneling rate, $\Gamma^{S(D)}_{bb}$ through the BS and $\Gamma^{S(D)}_{aa}$, through the AS from the source (drain) electrode to the DQD. They are described as \cite{Kubo} 
\begin{eqnarray}
\Gamma^{S(D)}_{bb}=\Gamma_{s(d)}\{1+\alpha_{s(d)}\cos(\pi\Phi/\Phi_0)\}
\label{eqn:rateb}\\
\Gamma^{S(D)}_{aa}=\Gamma_{s(d)}\{1-\alpha_{s(d)}\cos(\pi\Phi/\Phi_0)\}
\label{eqn:ratea}
\end{eqnarray}
where $\Gamma_{s(d)}$ corresponds to the tunneling rate from the source (drain) electrode to the DQD at $\alpha_s=\alpha_d=0$, otherwise the $\alpha_{s(d)}$ parameters are assumed to be the same between the BS and AS. 
Because the higher order tunneling processes comprising both the BS and AS can be neglected due to the small $\alpha_{s(d)}$ in our case as explained above, the currents, $I_b$ through the BS,  and $I_a$ through the AS are expressed as
\begin{eqnarray}
I_{b(a)}\propto(1/\Gamma^{S}_{bb(aa)}+1/\Gamma^{D}_{bb(aa)})^{-1}, 
\label{eqn:current}
\end{eqnarray}
respectively \cite{epaps}. When $\alpha_s$ and $\alpha_d$ have the same sign, $I_{b}$ ($I_{a}$) directly reflects the oscillatory parts in $\Gamma^{S}_{bb}$ and $\Gamma^{D}_{bb}$ ($\Gamma^{S}_{aa}$ and $\Gamma^{D}_{aa}$) with the same period of $2\Phi_0$ in phase. Then, as expected from Eqs.~(\ref{eqn:rateb}) and (\ref{eqn:ratea}), the AB oscillation phase is different by a half period, i.e. $\Phi_0$, between $I_{b}$ and $I_{a}$. 
On the other hand, when $\alpha_s$ and $\alpha_d$ have opposite signs, the oscillation amplitude of $I_{b}$ ($I_{a}$) becomes small because $\Gamma^{S}_{bb}$ and $\Gamma^{D}_{bb}$ ($\Gamma^{S}_{aa}$ and $\Gamma^{D}_{aa}$) oscillate with the same period of $2\Phi_0$ but in the opposite phase. 
Consequently, the AB oscillations of $I_{b}$ and $I_{a}$ appear to  be in phase with the period of $2\Phi_0$ but with an extra dip or peak, although the visibility of the AB oscillations should be lower than that when $\alpha_s$ and $\alpha_d$ have the same sign. 
Therefore, it is possible to account for the changes in the period and phase of the AB oscillation with respect to $V_c$ by using $\alpha_{s(d)}$. 

\begin{figure}
\centering
\includegraphics[width=1\columnwidth]{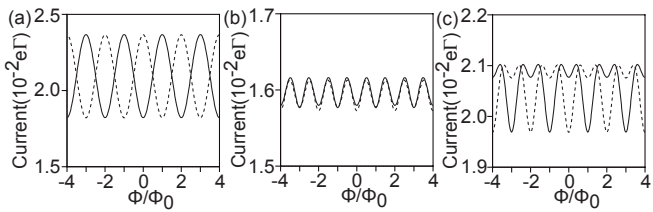}
\caption{
Currents flowing through the bonding (solid line) and anti-bonding (dashed line) states as a function of magnetic flux $\Phi$ at (a) $\alpha_s=-0.1$ and $\alpha_d=-0.2$, (b) $\alpha_s=-0.1$ and $\alpha_d=0.2$ and (c) $\alpha_s=-0.1$ and $\alpha_d=0.15$, where $\Gamma=\Gamma_s+\Gamma_d$. 
} 
\label{theory}
\end{figure}

Here, let us discuss the experimental result in Fig.~\ref{figure3} concretely, using the change of $\alpha_{s(d)}$. 
Initially, we consider the signs of $\alpha_s$ and $\alpha_d$ at $V_c=-1.20$ V in our experiment. 
The phases of the AB oscillations for the BS and the AS in Fig.~\ref{figure3}(a) and (b) are not consistent with those predicted for $\alpha_s=\alpha_d=1$ \cite{Kang}. This discrepancy can be explained by assuming that both $\alpha_s$ and $\alpha_d$ have negative signs. 
Then, we consider the changes in $\alpha_s$ and $\alpha_d$ as $V_c$ becomes more negative. $\alpha_{s(d)}$ decreases to 0 with oscillation  as the path length in the source (drain) electrode increases \cite{epaps}. 
$\alpha_d$ probably depends on the change in $V_c$ more strongly than $\alpha_s$, because the screening effect is stronger in the drain than in the source in the side of the substrate. 
Therefore, for simplicity we assume that $\alpha_s$ remains a negative constant as $\alpha_s=-0.1$, and that $\alpha_d$ changes from negative to positive, i.e. $\alpha_d=-0.2$ to 0.2 then to 0.15 and we calculate the currents flowing through the BS and AS numerically using the non-equilibrium Green's function method. The results are shown in Fig.~\ref{theory} (a) to (c). We used $2t/\hbar\Gamma=-265$ and $eV_{SD}/\hbar\Gamma =400$ in (a), $2t/\hbar\Gamma =-175$ and $eV_{SD}/\hbar\Gamma =250$ in (b), and $2t/\hbar\Gamma =-120$ and $eV_{SD}/\hbar\Gamma =250$ in (c), with $V_{inter}/\hbar\Gamma =200$, $k_BT/\hbar\Gamma=10$ as common parameters, where $k_B$ is the Boltzmann constant, $T$ is the temperature, $\hbar=h/2\pi$ and $\Gamma=\Gamma_s+\Gamma_d$ ($\Gamma_s=0.3\Gamma$ and $\Gamma_d=0.7\Gamma$), respectively. We evaluated all these parameters experimentally.

Figure \ref{theory}(a) shows AB oscillations with the same period of $2\Phi_0$ but misaligned by $\Phi_0$ between the BS and AS, so that the BS and AS currents have a dip, and a peak at $\Phi=0$, respectively. These features are consistent with those at $V_c=-1.20$ V in Fig.~\ref{figure3}(a) and (b). 
When $\alpha_d$ becomes 0.2, which corresponds to the more negative $V_c$ \cite{epaps}, 
Fig. \ref{theory}(b) shows almost the same AB oscillations with a period $\Phi_0$ for both BS and AS; the period of the AB oscillation changes. Furthermore, the AS current at $\Phi=0$ in (b) is a dip, while it is a peak in Fig.~\ref{theory}(a); the phase of the AB oscillation of the AS current also changes.  
These features are comparable to those in Fig.~\ref{figure3} (e) and (f) ($V_c=-1.36$ V). 
When $\alpha_d$ becomes 0.15, which corresponds to a more negative $V_c$, Fig. \ref{theory} (c) also shows AB oscillations with a period $2\Phi_0$ for both BS and AS but with an extra small dip.
This feature is similar to that for the AS current in Fig.~\ref{figure3} (h) at $V_c$=-1.40 V, although the AB oscillation for the BS current is not clear in Fig.~\ref{figure3}(g). 
Note that the features of the calculated AB oscillation change smoothly from Fig.~\ref{theory}(a) to (c) \cite{epaps}. 
Regarding Fig.~\ref{figure3}(c) and (d), because $V_c=-1.26$ V is an intermediate point between $V_c=-1.20$ and -1.36 V, $\alpha_d$ probably represents the change from negative to positive. Therefore, the AB oscillations of the BS and AS are not as clear as those at $V_c=-1.20$ V because of the small absolute value of $\alpha_d$. 
Note that the small dip in the AS current at $\Phi=0$ in Fig.~\ref{figure2}(d) probably evolves into the large dip in Fig.~\ref{figure2}(f). 

The AB oscillations of the BS are not as clear as those of the AS in Fig.~\ref{figure3}(e)-(h). We do not fully understand the reason for this, but consider that the $\Gamma_{s(d)}$ value of the BS is smaller than that of the AS at around $\Phi \approx 0$ when $\alpha_{s(d)}$ is negative (see Eq.(\ref{eqn:rateb}) and (\ref{eqn:ratea})), resulting in the smaller BS current at around zero magnetic field.

In conclusion, we have investigated AB oscillation for the BS and AS in a DQD. The periods of the AB oscillations are roughly halved and, for the AS, the AB phase shifts by $\Phi_0$, depending on the center gate voltage. 
These features are assigned to contributions from indirect interdot tunnel coupling, which depends on the center gate voltage. 

We thank Y. -S. Shin, A. Shibatomi, S. Sasaki and W. Izumida for experimental help and fruitful discussions. 
Part of this work is financially supported by JSPS Grant-in-Aid for Scientific Research S (No. 19104007), MEXT Grant-in-Aid for Scientific Research on Innovative Areas (21102003), Funding Program for World-Leading Innovative R\&D on Science and Technology(FIRST), and IARPA grant W911NS-10-1-0330.


\clearpage
\widetext
\setcounter{page}{1}
\setcounter{figure}{0}
\setcounter{equation}{0}

\renewcommand{\thepage}{S\arabic{page}}
\renewcommand{\thefigure}{S\arabic{figure}}

\begin{center}

{\Large
Supplementary Material for

\vspace{0.5cm}

\baselineskip 20mm
" Aharonov-Bohm Oscillations Changed by Indirect Interdot Tunneling via Electrodes in Parallel-Coupled Vertical Double Quantum Dots "
}

\vspace{0.5cm}

by

\vspace{0.3cm}

T. Hatano, T. Kubo, Y. Tokura, S. Amaha, S. Teraoka and S. Tarucha 
\end{center}

\section{Device structure}

\begin{figure}[hb]
\centering
\includegraphics[width=1\columnwidth]{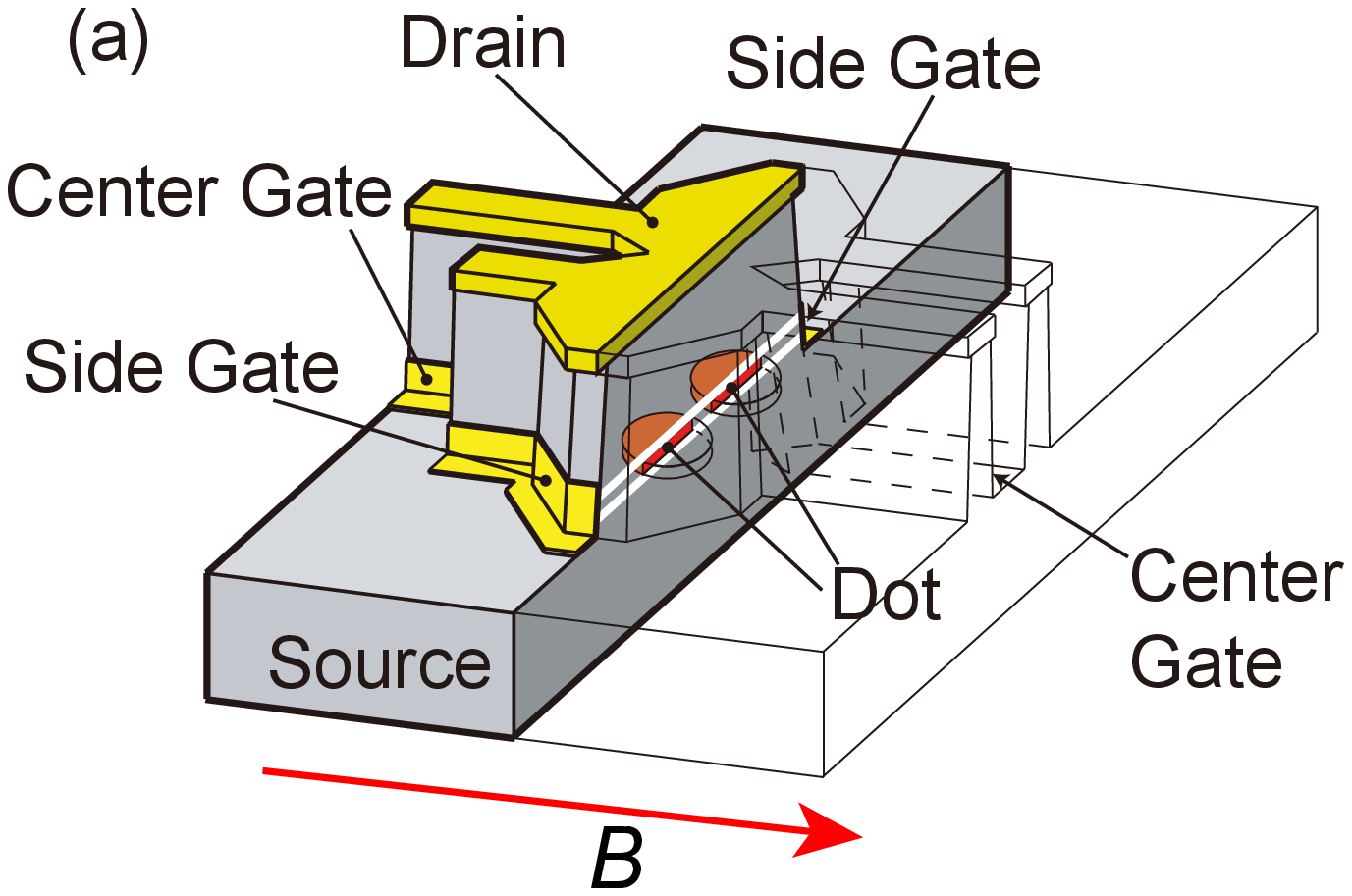}
\caption{
Schematic of double quantum dot device. The red arrow shows the direction of the applied magnetic field $B$. 
} 
\end{figure}

Our device consists of two laterally coupled vertical quantum dots (QDs) with four split gates made from a double-barrier heterostructure (DBH) of an undoped 10 nm GaAs well and two undoped AlGaAs barriers with thicknesses of 8.0 and 7.0 nm as shown in Fig.~S1(a)) \cite{austing,hatano}. 
Two of the four gates (side gates) are used to tune the electron number in each QD independently, and the remaining two gates (center gates) are used to modulate the strength of the interdot tunnel coupling. 
A current flows vertically through the two parallel coupled QDs between the common source and drain as shown in Fig.~1(b). 

The calculated area, $S/2$, of the two current loops formed by the double QD (DQD) and either the source or drain electrodes due to the presence of the direct interdot tunnel coupling is approximately 350 nm $\times$ 15 nm$ =5.3\times 10^{-15}$ m$^2$, using our estimates of $\sim$350 nm for the dot diameter and $\sim$15 nm for the half value of the length of the double barrier structure.

\section{Indirect coherent coupling parameter "$\alpha$"}

\begin{figure}[hbt]
\centering
\includegraphics[width=1\columnwidth]{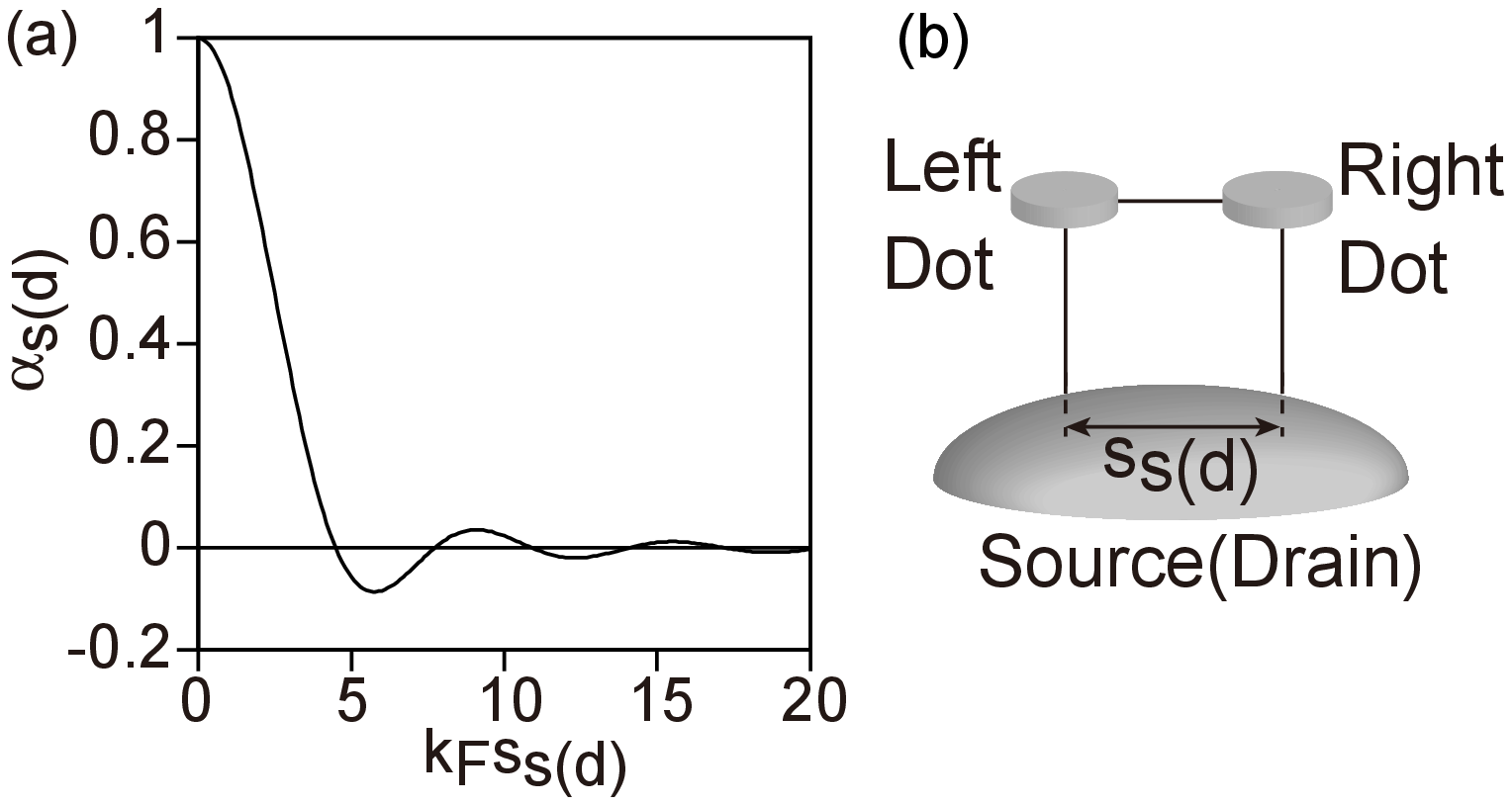}
\caption{
(a) Model of source (drain) electrode and two dots. $s_{s(d)}$ correspond to the distance between the two dots in the source (drain) electrode. (b)Dependence of $\alpha_{s(d)}$ on length between two dots $s_{s(d)}$. 
} 
\label{model}
\end{figure}

The indirect coupling parameter $\alpha$$(|\alpha| \leqq 1)$ represents the degree of coherence in the electrodes \cite{kubo}.
Figure~\ref{model} shows the calculation of $\alpha_{s(d)}$ as a function of the path length, $s_{s(d)}$, in the source (drain) electrode to connect the two dots, when we simply assume one-dimensional electron propagation over the length $s_{s(d)}$ as shown in Fig.~\ref{model}(b) \cite{kubo}. $k_F$ is the Fermi wave number. As shown in Fig.~\ref{model}(a), $\alpha_{s(d)}$ decreases to 0 with an oscillation as $s_{s(d)}$ increases. 
Note that the $s_{s(d)}$ dependence of $\alpha_{s(d)}$ in (a) is stronger  than that assumed in the calculation in Fig.~3. This is probably because the model (b) used for the calculation is too simple to provide a quantitative estimation of $\alpha_{s(d)}$ in the experiment. 

\section{Derivation of tunneling current through a DQD}

We derive the tunneling current through the DQD. 
For clarity of discussion, we concentrate on the noninteracting situation, and neglect the spin degree of freedom. In the last part of this section, the effect of the interdot Coulomb interaction is taken into account numerically.

Using the nonequilibrium Green's function method, we obtain the following expression for the tunneling current through a DQD \cite{current}
\begin{eqnarray}
I_{DQD}(\phi)=\frac{e}{2\pi}\int\frac{d\epsilon}{\hbar}[f_S(\epsilon)-f_D(\epsilon)]T(\epsilon,\phi).\label{current}
\end{eqnarray}
Here $f_{\nu}(\epsilon)$ is the Fermi-Dirac distribution function of the reservoir $\nu$ defined by
\begin{eqnarray}
f_{\nu}(\epsilon)=\frac{1}{e^{(\epsilon-\mu_{\nu})/k_BT}+1},
\end{eqnarray}
where $\mu_{\nu}$ is the electrochemical potential of the reservoir $\nu$, and $T(\epsilon,\phi)$ is the transmission probability defined as
\begin{eqnarray}
T(\epsilon,\phi)&\equiv&\mbox{Tr}\left\{\bm{G}^r(\epsilon,\phi)\bm{\Gamma}^S(\phi)\bm{G}^a(\epsilon,\phi)\bm{\Gamma}^D(\phi) \right\},
\end{eqnarray}
where the boldface notation indicates a $2\times 2$ matrix whose basis is the tunnel-coupled bonding ($b$) and antibonding ($a$) states in a DQD. $\bm{G}^{r(a)}(\epsilon,\phi)$ is the retarded (advanced) Green's function matrix and $\bm{\Gamma}^{\nu}(\phi)$ is the linewidth function matrix for the reservoir $\nu\in S,D$ given by
\begin{equation}
\bm{G}^r(\epsilon,\phi)=\left(
  \begin{array}{cc}
     \frac{\epsilon-\epsilon_b}{\hbar}+\frac{i}{2}\Gamma_{bb}(\phi)      &  \frac{i}{2}\Gamma_{ba}(\phi)  \\
     \frac{i}{2}\Gamma_{ab}(\phi)      &  \frac{\epsilon-\epsilon_a}{\hbar}+\frac{i}{2}\Gamma_{aa}(\phi)  \\
  \end{array}
\right)^{-1},
\end{equation}
\begin{equation}
\bm{\Gamma}^{S}(\phi)=\Gamma_s\left(
  \begin{array}{cc}
     1+\alpha_s\cos(\pi\Phi/\Phi_0)      & i\alpha_s\sin(\pi\Phi/\Phi_0)   \\
     -i\alpha_s\sin(\pi\Phi/\Phi_0)      & 1-\alpha_s\cos(\pi\Phi/\Phi_0)   \\
  \end{array}
\right)\label{lws}
\end{equation}
\begin{equation}
\bm{\Gamma}^{D}(\phi)=\Gamma_d\left(
  \begin{array}{cc}
     1+\alpha_d\cos(\pi\Phi/\Phi_0)      & -i\alpha_d\sin(\pi\Phi/\Phi_0)   \\
     i\alpha_d\sin(\pi\Phi/\Phi_0)      & 1-\alpha_d\cos(\pi\Phi/\Phi_0)   \\
  \end{array}
\right),\label{lwd}
\end{equation}
\begin{equation}
\bm{\Gamma}(\phi)=\bm{\Gamma}^{S}(\phi)+\bm{\Gamma}^{D}(\phi).
\end{equation}
Here $\epsilon_{b(a)}$ is the energy of the tunnel-coupled bonding and antibonding states, and $\alpha_{\nu}$ is the coherent indirect coupling for the reservoir $\nu$ \cite{kubo}, and $\phi$ is the AB phase defined as $\phi=2\pi\Phi/\Phi_0$, where $\Phi$ is the magnetic flux threading through an AB interferometer and $\Phi_0=h/e$ is the magnetic flux quantum. Moreover, we introduce the notation $M_{ij}$, which denotes the $(i,j)$ matrix element of $2\times 2$ matrix $\bm{M}$.

Using the above results, we obtain the following expressions for the tunneling current
\begin{eqnarray}
I_{SD}(\phi)&=&I_b(\phi)+I_a(\phi)+I_{ba}(\phi),\\
I_b(\phi)&=&\frac{e}{2\pi}\int\frac{d\epsilon}{\hbar}[f_S(\epsilon)-f_D(\epsilon)]\Gamma_{bb}^S\Gamma_{bb}^D|G_{bb}^r(\epsilon,\phi)|^2,\\
I_a(\phi)&=&\frac{e}{2\pi}\int\frac{d\epsilon}{\hbar}[f_S(\epsilon)-f_D(\epsilon)]\Gamma_{aa}^S\Gamma_{aa}^D|G_{aa}^r(\epsilon,\phi)|^2,\\
I_{ba}(\phi)&=&\frac{e}{2\pi}\int\frac{d\epsilon}{\hbar}[f_S(\epsilon)-f_D(\epsilon)]\left[\Gamma_{bb}^S\Gamma_{aa}^D|G_{ab}^r(\epsilon,\phi)|^2+\Gamma_{aa}^S\Gamma_{bb}^D|G_{ba}^r(\epsilon,\phi)|^2 \right.\nonumber\\
&&+2\mbox{Re}\left\{\Gamma_{bb}^S\Gamma_{ab}^DG_{bb}^r(\epsilon,\phi)[G_{ab}^r(\epsilon,\phi)]^*+\Gamma_{ba}^S\Gamma_{bb}^DG_{bb}^r(\epsilon,\phi)[G_{ba}^r(\epsilon,\phi)]^* \right.\nonumber\\
&&+\Gamma_{ab}^S\Gamma_{ab}^DG_{ba}^r(\epsilon,\phi)[G_{ab}^r(\epsilon,\phi)]^*+\Gamma_{ba}^S\Gamma_{ab}^DG_{bb}^r(\epsilon,\phi)[G_{aa}^r(\epsilon,\phi)]^*\nonumber\\
&&\left. \left.
+\Gamma_{aa}^S\Gamma_{ab}^DG_{ba}^r(\epsilon,\phi)[G_{aa}^r(\epsilon)]^*+\Gamma_{ba}^S\Gamma_{aa}^DG_{ab}^r(\epsilon,\phi)[G_{aa}^r(\epsilon,\phi)]^* \right\}\right].
\end{eqnarray}
Here we evaluate $I_b(\phi)$ in a high bias voltage regime such as $eV_{SD}\gg \epsilon_b-\epsilon_a=2t_c, \Gamma_{ij}^{\nu}(\phi), k_BT$
\begin{eqnarray}
I_b(\phi)&=&e\frac{\Gamma_{bb}^S\Gamma_{bb}^D}{\Gamma_{bb}}\left[1+O(\alpha_{\nu}\alpha_{\nu'}) \right],
\end{eqnarray}
where $t_c$ is the direct interdot tunnel coupling. In our paper, we focus on \textit{weak} coherent indirect coupling conditions, namely $|\alpha_{\nu}|\ll 1$. We assume that the terms proportional to $\alpha_{\nu}\alpha_{\nu'}$ are negligible. Then, we have
\begin{eqnarray}
I_b(\phi)&\simeq&e\frac{\Gamma_{bb}^S\Gamma_{bb}^D}{\Gamma_{bb}}\nonumber\\
&=&e\frac{1}{\frac{1}{\Gamma_{bb}^S}+\frac{1}{\Gamma_{bb}^D}}.
\end{eqnarray}
This is equivalent to Eq. (3) in the main text. This result has the form of a Breit-Wigner type transmission through the bonding state. Therefore, we call $I_b(\phi)$ the bonding state component of the tunneling current. Similarly, we can derive the expression of the antibonding state component of the tunneling current.
\begin{eqnarray}
I_a(\phi)\simeq e\frac{1}{\frac{1}{\Gamma_{aa}^S}+\frac{1}{\Gamma_{aa}^D}}.
\end{eqnarray}
Moreover, the mixing component of the bonding and antibonding states for the tunneling current is given as
\begin{eqnarray}
I_{ba}(\phi)\propto O(\alpha_{\nu}\alpha_{\nu'}).
\end{eqnarray}
Thus, this mixing component is negligible when we use the above approximation.

\begin{figure}
\centering
\includegraphics[width=1\columnwidth]{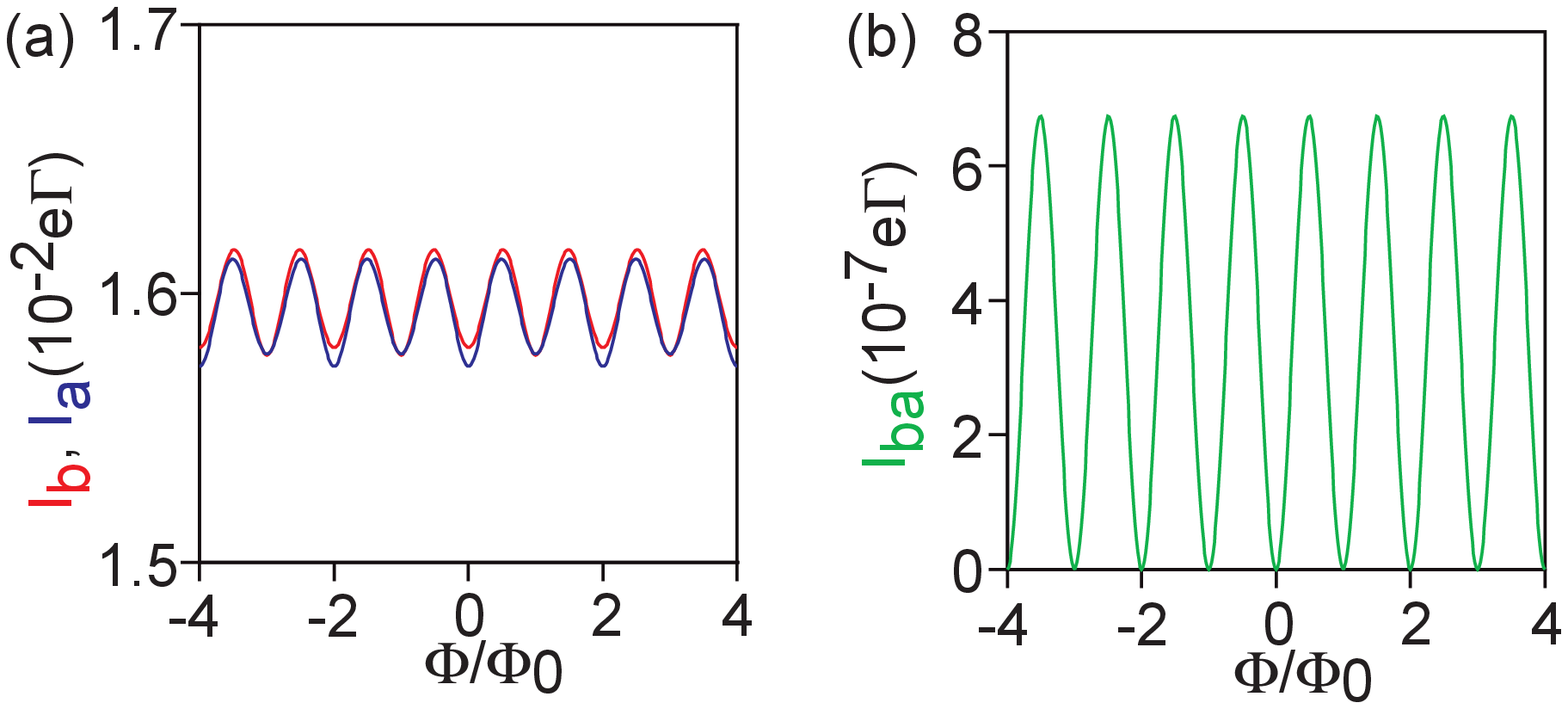}
\caption{Numerical results for the tunneling current. (a) Bonding and antibonding components. (b) Mixing component.}
\label{interaction} 
\end{figure}

The above results are the same even when there is a finite interdot Coulomb interaction. Under the condition shown in Fig. 3 (b) in the main text, we provide numerical results for the bonding, antibonding, and mixing components for the tunneling current in Fig. \ref{interaction}. It is clear that the mixing component $I_{ba}(\phi)$ of the tunneling current is much smaller than the bonding $I_b(\phi)$ and antibonding $I_a(\phi)$ components of the tunneling current. Moreover, we find that the mixing component vanishes when $\Phi/\Phi_0$ is an integer. This can be explained as follows. According to Eqs. (\ref{lws}) and (\ref{lwd}), the off-diagonal matrix elements of the linewidth functions are zero when $\Phi/\Phi_0$ is an integer. In this case, the bonding state is orthogonal to the antibonding state. Thus, the bonding and antibonding states do not contribute mixing component, namely $I_{ba}(\phi)=0$.

\section{$\alpha_s$ and $\alpha_d$ dependences of AB oscillation of bonding and antibonding states}

\begin{figure}[hbt]
\centering
\includegraphics[width=1\columnwidth]{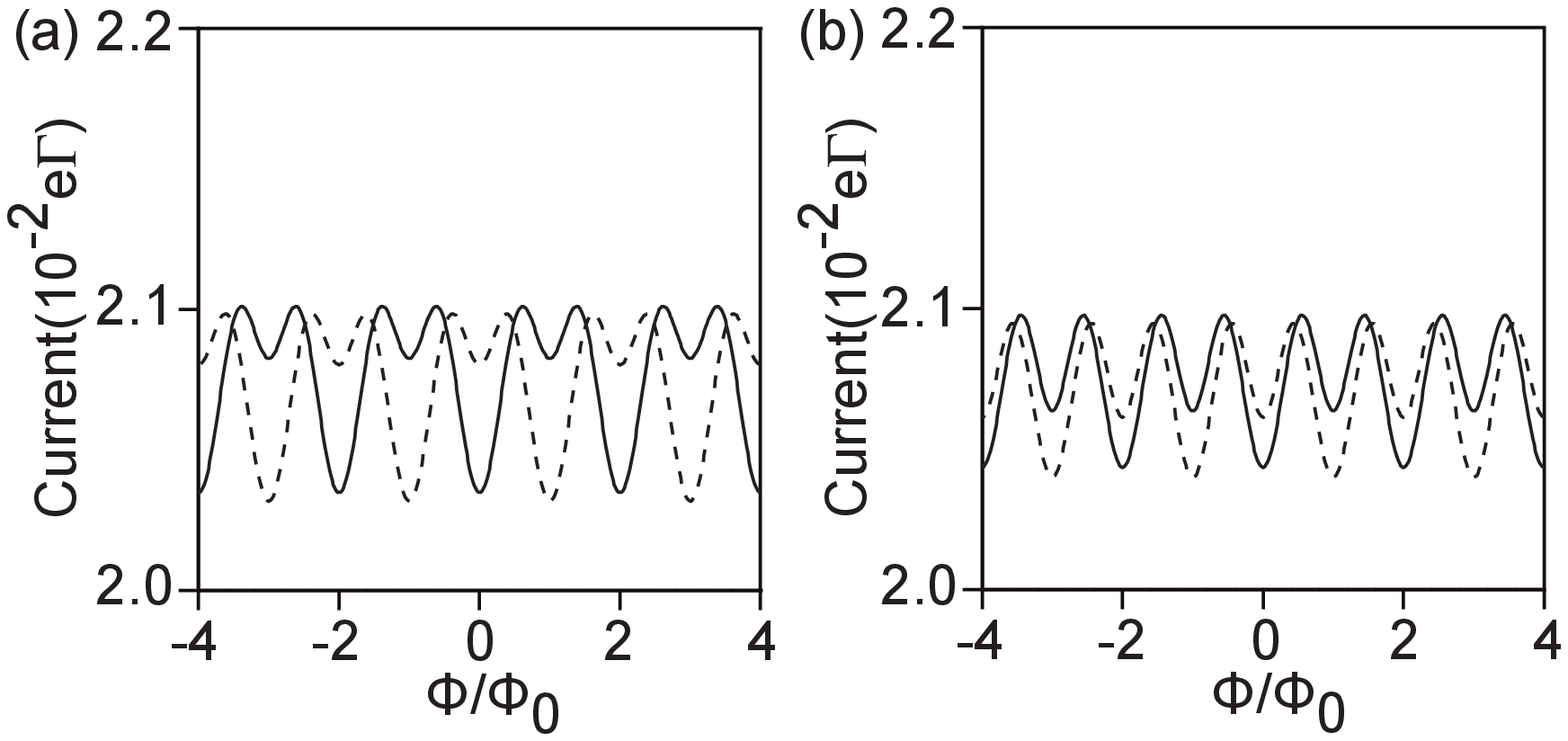}
\caption{
Currents through the bonding (solid line) and antibonding (dashed line) states as a function of a magnetic flux $\Phi$, for the indirect coupling parameter in the source and drain electrode, (a) $\alpha_s=-0.1$ and $\alpha_d=0.14$ and (b) $\alpha_s=-0.1$ and $\alpha_d=0.18$} 
\label{theory2}
\end{figure}

We calculate the bonding and antibonding state currents at an intermediate $\alpha_d$ value in Fig.~\ref{theory2}(a) between Fig.~3(a) and (b), and in Fig.~\ref{theory2}(b) between (b) and(c), respectively. 
We used $2t/\hbar\Gamma=-200$ and $eV_{SD}/\hbar\Gamma =250$ in (a) and $2t/\hbar\Gamma =-150$ and $eV_{SD}/\hbar\Gamma =250$ in (b), respectively, with $V_{inter}/\hbar\Gamma =200$ and $k_BT/\hbar\Gamma =10$ as the common parameters.

In Fig.~3(a), the period of the oscillations for the bonding and antibonding states is $2\Phi_0$ with a dip at $\Phi=2n\Phi_0$ ($\Phi =(2n+1)\Phi_0$) for the bonding (antibonding) state, where $n$ is an integer. As $\alpha_d$ increases to $\sim 0.14$, the oscillation for the bonding (antibonding) state has an additional small dip at $\Phi=(2n+1)\Phi_0$ ($\Phi=2n\Phi_0$) as shown in Fig.~\ref{theory2}(a). Thus, the period of the AB oscillations for the bonding and antibonding states changes to $\Phi_0$ but with two alternating amplitudes. 
The small dip grows and becomes comparable to the other dip when $\alpha_d\approx0.2$ as shown in Fig.~3(b). 
As $\alpha_d$ decreases to $\sim 0.18$, the difference between the two alternating oscillation amplitudes increases again (see Fig.~\ref{theory2} (b)), and finally, the AB oscillations for the bonding and antibonding states evolve into those seen in Fig.~3(c).

\section{Aharonov-Bohm oscillation for two-electron ground state}

\begin{figure}
\centering
\includegraphics[width=1\columnwidth]{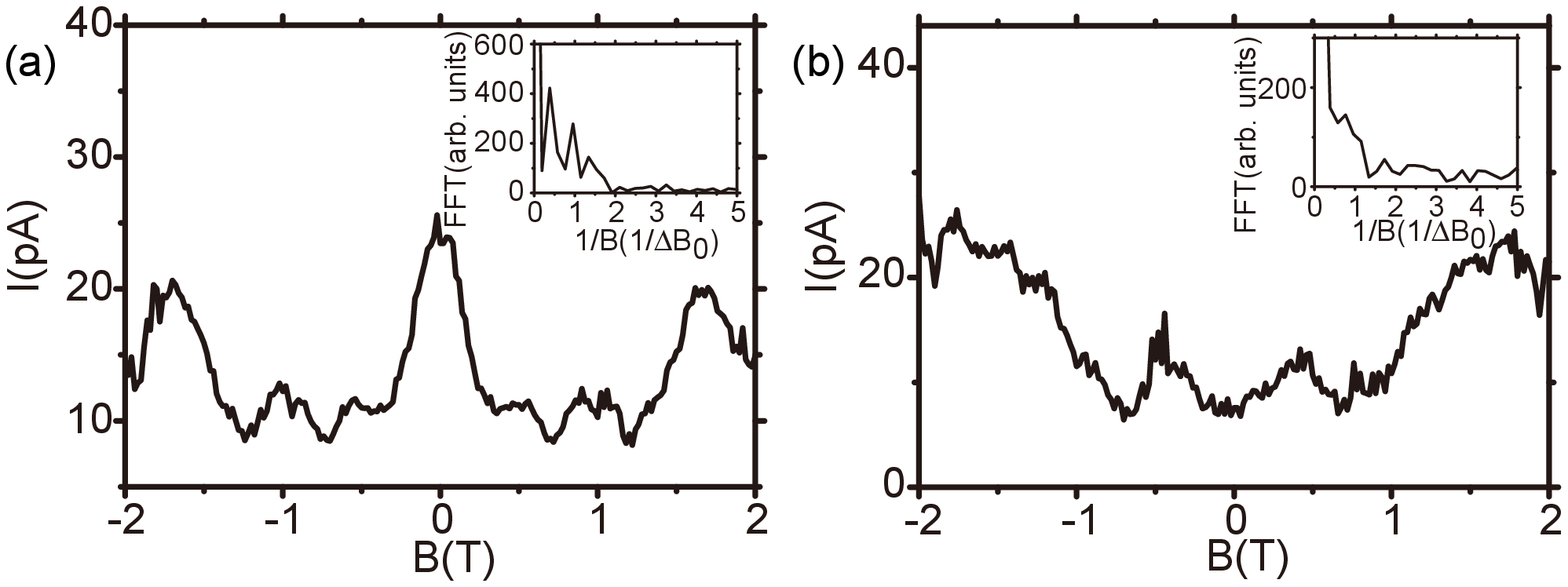}
\caption{
Current flowing through the two-electron state at (a) $V_c=$-1.20 V and (b) $V_c=$-1.36 V as a function of $B$. The insets show the FFT spectra of the current oscillations. 
} 
\label{2electrons}
\end{figure}

The coherence, i.e. the AB oscillation, of the two-electron state is attracting interest to detect the singlet and triplet states \cite{loss}, similarly to the that of the one-electron states.  
Consequently, we investigate the AB oscillation for the two-electron ground state. Figure \ref{2electrons}(a) and (b) show experimental data obtained at $V_{c}=$-1.20  and -1.36 V, respectively. 
We observe periodic current oscillations with a large peak at $B=0$ T in (a) but no clear oscillations in (b). The FFT spectrum in (a) indicates two oscillation periods of 0.81 T and 1.8 T. For the first period, we find the same period in Fig.~2(a) and (b) but not for the other, which arises from the three large peaks at 0 and $\pm$1.8 T in Fig.~\ref{2electrons}(a). 

We used a master equation method expanded from that reported in \cite{tokura2} to calculate the current flowing through the two-electron ground state.

When the exchange coupling energy is very small the singlet and triplet states are not well resolved. 
Then, the Heitler-London (HL) state comprising equivalent contributions from the bonding and anti-bonding states is the two-electron ground state. The current through the HL state does not oscillate with the magnetic field because the oscillations of the bonding and antibonding states are canceled out \cite{tokura2}. 
Therefore, we can obtain the AB oscillation in the relatively strong tunnel coupling case as shown in Fig.\ref{2electrons}(a). 

On the other hand, when the exchange coupling is strong enough for the singlet and triplet states to be well separated, the singlet state is composed of the HL and molecular states for which two electrons with anti-parallel spins occupy the bonding state. Therefore, the current through the molecular state oscillates. 
Accordingly, we cannot obtain the AB oscillation in the weak tunnel coupling case as shown in Fig.\ref{2electrons}(b). 

The two phases of the AB oscillations in Figs.~\ref{2electrons} (a) and (b) are the same as those of the anti-bonding state as shown in Figs.~2(b) and(f), respectively. However, this is not consistent with the theory, because the AB oscillation phase of the molecular state is the same as that of the bonding state \cite{tokura2}. This discrepancy can be understood if $\alpha_s$ and $\alpha_d$ for the two-electron state are different from those for the one-electron state. 
This argument is reasonable, because the $V_{sL}$ and $V_{sR}$ values of the one-electron and two-electron states are different, even when $V_c$ is the same.
A more detailed study is necessary for the two electron ground state.



\end{document}